\documentclass[onecolumn,nobibnotes,nofootinbib,superscriptaddress]{revtex4}
\usepackage{amsmath,amssymb}
\usepackage{graphicx,subfigure,epsfig}

\newcommand{\be}{\begin{equation}}
\newcommand{\ee}{\end{equation}}
\newcommand{\bea}{\begin{eqnarray}}
\newcommand{\eea}{\end{eqnarray}}
\newcommand{\benn}{\begin{displaymath}}
\newcommand{\eenn}{\end{displaymath}}
\newcommand{\beann}{\begin{eqnarray*}}
\newcommand{\eeann}{\end{eqnarray*}}
\newcommand{\oone}{
\begin{picture}(10,8)
\put(5,5){\circle{8}}
\put(2.9,2.5){{\scriptsize 1}}
\end{picture}
}

\newcommand{\beq}{\begin{equation}}
\newcommand{\eeq}{\end{equation}}
\newcommand{\nn}{\nonumber\\}
\newcommand{\dif}{{\rm d}}

\newcommand{\bx}{x_{\bot}}
\newcommand{\by}{y_{\bot}}
\newcommand{\bu}{z'_{\bot}}

\newcommand{\bz}{z_{\bot}}

\newcommand{\abar}{\bar{\alpha}_s}

\newcommand{\Nc}{N_{\rm c}}

\newcommand{\Nf}{N_{\rm f}}

\newcommand{\minus}{\!-\!}

\begin{document}

\title{Rare fluctuations of the $S$-matrix at NLO in QCD}
\author{Wenchang Xiang}
\email{wxiangphy@gmail.com}
\affiliation{Guizhou Key Laboratory in Physics and Related Areas, Guizhou University of Finance and Economics, Guiyang 550025, China}
\affiliation{Department of Physics, Guizhou University, Guiyang 550025, China}
\author{Yanbing Cai}
\email{myparticle@163.com}
\affiliation{Guizhou Key Laboratory in Physics and Related Areas, Guizhou University of Finance and Economics, Guiyang 550025, China}
\author{Mengliang Wang}
\email{mengliang.wang@mail.gufe.edu.cn}
\affiliation{Guizhou Key Laboratory in Physics and Related Areas, Guizhou University of Finance and Economics, Guiyang 550025, China}
\author{Daicui Zhou}
\email{dczhou@mail.ccnu.edu.cn}
\affiliation{Key Laboratory of Quark and Lepton Physics (MOE) and Institute of Particle Physics, Central China Normal University, Wuhan 430079, China}

%\pacs{}

\begin{abstract}
We calculate the rare fluctuations of the $S$-matrix on top of the full next-to-leading order corrections in the center of mass frame. The relevant result in the saturation regime shows that the exponential factor of the $S$-matrix is $\sqrt{2}$ as large as the result which emerges when the rare fluctuation effects are taken into account. We find that the factor of $\sqrt{2}$ change of the exponential factor is induced by the gluon loop corrections which compensate part of rapidity decrease of the $S$-matrix made by quark loops and lead to the rare fluctuations becoming important again. To ensure the relevant results of the $S$-matrix are independent of the frame choice, the rare fluctuations of the $S$-matrix are also derived in a general frame. It is found that all the results are consistent with each other in both frames.
\end{abstract}

\maketitle

%-----------------------------------------------------------------------------

\section{Introduction}
\label{sec:intro}
In high energy QCD one of the most challenging problem has been to study theoretically and experimentally the parton saturation. This saturation phenomenon was firstly introduced by Gribov, Levin and Ryskin in 1983, and was developed as a dual description of unitarity\cite{Gribov}. Since then many physicists are engaged in searching of QCD evolution equations for describing the evolution of high energy and density gluon systems. Among them, the most widely used one is the Balitsky-Kovchegov (BK) equation\cite{Balitsky,Kovchegov} because of its relatively simple structure. The BK equation is included the non-linear effect which ensures the scattering amplitude fully satisfying the unitarity constraints instead of an exponential growth with rapidity in the BFKL dynamics\cite{BFKL1,BFKL2}. At leading logarithmic accuracy where the BK equation resums large logarithms $ \alpha_s\ln(1/x)$ corrections with fixed coupling constant $\alpha_s$, the analytical\cite{Mueller-Triantafy,IIM} and numerical\cite{Albacete} studies of the BK equation show that the BK equation with only leading order (LO) contribution cannot precisely describe the high energy scattering in small-$x$ region.

In the past decade, the remarkable developments beyond leading logarithmic accuracy have been made to systematically improve the precision of the BK equation. One of the most important efforts was the consideration of the running coupling corrections. The running coupling BK (rcBK) equations were independently derived by Balitsky in Ref.\cite{Balitskyrc} and Kovchegov-Weigert in Ref.\cite{Kovchegov-Weigertrc} via resummation all order corrections associated with the coupling. However, the rcBK equations resulting from these two groups have different format due to using different separation schemes between the running coupling and subtraction. Fortunately, in the saturation region (the most interesting regime in this paper) our studies in Ref.\cite{Xiang} show that these two rcBK equations reduce to an uniform format and have a same solution. In a full region, the numerical solutions to the rcBK equations show that the evolution speed of the scattering amplitude from Balitsky's derivation is dramatically suppressed as compared to the one from the fixed coupling BK equation\cite{Albacete-Kovchegov}, which is coincide with the theoretical expectations. %While the solutions of Kovchegov-Weigert's equation for different rapidities are close to the ones from the fixed coupling BK equation, which is far from theoretical expectations.
Although the running coupling corrections significantly slow down the evolution speed of the dipole amplitude as rapidity increasing, they are not the only higher order corrections relative to LO BK equation. Indeed in the language of Feynman diagram the running coupling corrections are only the contribution from quark loops, as we know that the gluon loops also have a large contribution to the evolution kernel\cite{Balitsky-Chirilli}. By combining the contributions from quark and gluon loops together, Balitsky and Chirilli got a full next-to-leading order (NLO) BK equation\cite{Balitsky-Chirilli}. The first numerical solution of the full NLO BK equation shows that the solution is sensitive to the details of the initial condition and becomes negative and non-vanishing at very small dipole size, which is unphysical\cite{Lappi-Mantysaari}. The origin of the instability of the solution is due to evolution equation including a large double transverse logarithmic correction term
($\sim \ln[(x_{\bot}-z_{\bot})^2/(x_{\bot}-y_{\bot})^2]\ln[(y_{\bot}-z_{\bot})^2/(x_{\bot}-y_{\bot})^2]$). Fortunately, the studies in Ref.\cite{IMMST} found that the instability problem can be solved by a resummation scheme for the double transverse logarithms.

Parallel to the developments in the calculations of contributions from quark and gluon loops, another effort in higher order contributions is also performed by including the effect of rare fluctuations to improve the precision of the BK equation\cite{Iancu-Mueller,Mueller-Shoshi,Xiang,Mueller-Munier,Liou,Laura}. As discussed by Iancu and Mueller in Ref.\cite{Iancu-Mueller}, a typical configuration includes too many gluons at the time of collision, therefore leading to a very small $S$-matrix. A rare configuration containing small number of gluons in the wavefunction has been found by suppressing the evolution of the gluons, which can lead to a relatively large $S$-matrix. In the fixed coupling case, the $S$-matrix obtained from BK equation shows a quadratic rapidity dependence on its exponent, and it was found that the exponential factor of the $S-$matrix ($\sim\exp{[-c_1\bar{\alpha}_s^2(Y-Y_0)^2/2]}$) is twice as large as the one which takes into account the rare fluctuation effects\cite{Iancu-Mueller}, where $\bar{\alpha}_s$ is the coupling constant and $c_1$ is a constant which is not important in this paper. This result shows that the rare fluctuations reduce the evolution speed of the dipole scattering amplitude with respect to rapidity. We would like to note that the result is obtained by including the rare fluctuations on top of the fixed coupling. How about the result of the rare fluctuations on top of the running coupling?

From the above discussions, we know that both running coupling and rare fluctuation effects can lead to suppress the evolution speed of the dipole scattering amplitude. To reveal which one is the dominant effect, we studied the rare fluctuations on top of running coupling effect in Ref.\cite{Xiang}. Firstly, we solved the rcBK equation in the saturation region to get an analytic $S$-matrix ($\sim\exp{[-N_c\mu(Y-Y_0)/\pi\mu_1]}$ with $\mu$ and $\mu_1$ coming from running coupling at one-loop accuracy) which shows a linear rapidity dependence in its exponent. Then the rare fluctuations is computed on top of this $S$-matrix, it was found that the rare fluctuation effects take a negligible change of the exponent of $S$-matrix, which mean that the rare fluctuations are less important in the running coupling case as compared to the fixed coupling case.

Recently, the solution to the full NLO BK equation is derived in the saturation region\cite{Xiang2}. The analytic result of the $S$-matrix shows that the $\exp{(-\mathcal{O}(Y))}$ rapidity dependence of the running coupling solution is replaced by $\exp{(-\mathcal{O}(Y^{3/2}))}$ in the NLO solution. As we know that the rare fluctuations in the fixed coupling case shows a significant suppression of the dipole amplitude, we believe that the rare fluctuations in full NLO case would also play an important role in the evolution of dipole amplitude, although the rare fluctuations are less important in the running coupling case.

In this paper, we investigate the rare fluctuations of the $S$-matrix on top of full NLO corrections in the saturation region. To see the rare fluctuation effects, we firstly recall the analytic solutions of LO, running coupling and next-to-leading order BK equations in the saturation region, and then study the rare fluctuation effects on top of these solutions. We find an interesting result that the exponential factor of the $S$-matrix from the NLO BK equation without rare fluctuations is about $\sqrt{2}$ times larger than the one with rare fluctuations.
%, which does not like the case in the rcBK equation where the rare fluctuation effects almost take no effect on the $S$-matrix.
The result shows that the influence of the rare fluctuations in the NLO BK case on the $S$-matrix are greater than that in the running coupling BK case, which indicate the rare fluctuations are important in the NLO BK case, although it is not as significant as the LO BK case.
%------------------------------------------------------------------------------
%\section{The LO, running coupling, NLL BK equations and their solutions in the saturation region}
%\label{sec:BK_eq}
%In this paper the Iancu-Mueller's approach for including the rare fluctuation effects is employed\cite{Iancu-Mueller}. In term's Iancu-Mueller's method, one needs to obtain the analytic expression of %$S$-matrix. In this section we briefly recall the LO, running coupling and NLL BK equations, and analytically solve them to get the $S$-matrixes in the saturation region.
%-----------------------------------------------------------
\section{Leading order evolution equation}
As we know that the simplest way to describe the scattering of a quark-antiquark dipole on a target (maybe another dipole, hadron or nucleus) in the high-energy regime is the BK equation which is a mean field version of the Balitsky-JIMWLK\footnote{The JIMWLK is the abbreviation of Jalilian-Marian, Iancu, McLerran, Weigert, Leonidov, Kovner.} hierarchy\cite{JIMWLK1,JIMWLK2,JIMWLK3,JIMWLK4} equations. The BK equation resums leading logarithmic $\alpha_s\ln(1/x)$ corrections with fixed coupling constant $\alpha_s$, which is a leading-order equation.

\subsection{Balitsky-Kovchegov equation}
Consider a scattering of a quark-antiquark dipole with a quark at transverse coordinate $x_{\bot}$ and an antiquark at transverse coordinate $y_{\bot}$ on a target, the dipole is left moving (unevolved), and the target is right moving (highly evolved), we usually call this frame as dipole frame\cite{Mueller}. In this frame almost all of the relative rapidity between dipole and target, $Y$, is taken by the target. If one increases the rapidity of the dipole by a small amount $dY$ while keeping the rapidity of the target fixed, then the dipole has a probability to emit a gluon at transverse coordinate $z_{\bot}$ due to the rapidity change. In the large $N_c$ limit the quark-antiquark-gluon state can be viewed as a system of two dipoles, which means a parent dipole splitting into two daughter dipoles. In the fixed coupling case, this evolution can be described by following equation\cite{Mueller}
\be
\frac{\partial}{\partial Y} S(r, Y) =
           \int d^2r_1 K^{LO}(r, r_1,r_2)
         \left [S^{(2)}(r_1, r_2,Y) -S(r, Y) \right ],
\label{eq_LOBK}
\ee
where the evolution kernel is given by
\be
K^{LO}(r, r_1, r_2)=\frac{\bar{\alpha}_s}{2 \pi} \frac{r^2}{r_1^2r_2^2} ,
\label{eq_LOBK_K}
\ee
with $\bar{\alpha}_s = \alpha_sN_c/\pi$. Here we use the notation ${\bf r}=x_{\bot}-y_{\bot}$ as the transverse size of parent dipole and ${\bf r}_1=x_{\bot}-z_{\bot}$ and ${\bf r}_2=z_{\bot}-y_{\bot}$ as the transverse sizes of the two emitted daughter dipoles, respectively. It is easy to see that the Eq.(\ref{eq_LOBK}) has a non-linear term on the right hand side, which accounts for the simultaneously scattering of the two daughter dipoles on the target. Equation (\ref{eq_LOBK}) is almost impossible to use directly since a solution for $S(r, Y)$ desires knowing $S^{(2)}(r_1, r_2,Y)$. In the the mean field approximation, one can simplify the non-linear term as following
\be
S^{(2)}(r_1, r_2, Y) \simeq
  S(r_1, Y) S(r_2, Y).
\label{mfa}
\ee
Substituting Eq.(\ref{mfa}) into Eq.(\ref{eq_LOBK}), one obtains the BK equation
\be
\frac{\partial}{\partial Y} S(r, Y) =
           \int d^2r_1 K^{LO}(r, r_1,r_2)
         \left [S(r_1, Y) S(r_2, Y) - S(r, Y) \right],
\label{LOBK}
\ee
which is a closed equation and can be solved analytically in the saturation region.

\subsection{Analytic solution in the saturation region}

In the saturation region where the unitarity corrections become important or $S$ is very small, the non-linear term in Eq.(\ref{LOBK}) is much smaller than the linear term. Therefore, the non-linear term can be neglected, Eq.(\ref{LOBK}) simplifies to
\be
\frac{\partial}{\partial Y} S(r, Y) =
           -\int d^2r_1 K^{LO}(r, r_1, r_2)
           S(r, Y).
\label{eq_LOBK_ap}
\ee
Since the dipole size is much larger than the characteristic size $1/Q_s$ in the saturation regime, the lower bound of integration in Eq.(\ref{eq_LOBK_ap}) can be set to $1/Q_s$, where $Q_s$ is the saturation scale. We set the upper bound of the integration to $r$ due to a rapid decrease of the integration beyond $r$. By analyzing the kernel of the integration in Eq.(\ref{eq_LOBK_ap}), one can know that the integration is dominated by the region either from $1/Q_s\ll \mid\mathbf{r_1}\mid\ll \mid\mathbf{r}\mid,\mid\mathbf{r_2}\mid\sim \mid\mathbf{r}\mid$ or from $1/Q_s\ll \mid\mathbf{r_2}\mid\ll \mid\mathbf{r}\mid,\mid\mathbf{r_1}\mid\sim \mid\mathbf{r}\mid$. Suppose we work in the region $\mid\mathbf{r_2}\mid\sim \mid\mathbf{r}\mid$, the kernel in Eq.(\ref{eq_LOBK_ap}) becomes $1/r_1^2$, which significantly simplifies the calculations. Now it is easy to get the analytic solution of the BK equation by performing the integrations over $r_1$ and $Y$\cite{Levin-Tuchin,Mueller}
\be
S(r, Y)=\exp\left[-\frac{c}{2}\bar{\alpha}_s^2(Y-Y_0)^2\right]S(r_0, Y_0),
\label{Sol_Kovchegov}
\ee
where we have used $Q_s^2(Y)=\exp\left[c\bar{\alpha}_s(Y-Y_0)\right]Q_s^2(Y_0)$ with $Q_s^2(Y_0)r^2=1$. From Eq.(\ref{Sol_Kovchegov}) we can see that the analytic solution of LO BK equation has a quadratic rapidity dependence in the exponent. However, we will see in the following sections that the solution will be modified by taking into account the higher-order contributions, especially rare fluctuation effects. The reason why we have gone through such a detailed "derivation" of Eq.(\ref{Sol_Kovchegov}) is that the major aim of this study is to show how the $S$-matrix is modified by rare fluctuation effects in the cases of LO, running coupling and NLO cases, respectively.
%-----------------------------------------------------------
\section{Next-to-leading order evolution equation}
\label{subsec_rcBK}
The LO BK equation discussed above considers only the resummation of leading logarithmic $\alpha_s\ln(1/x_{Bj})$ corrections with a fixed coupling constant. Beyond the leading logarithmic approximation, a significant progress in the evolution equation has been made via resummation of $\alpha_s N_f$ to all order which is usually called as the running coupling corrections\cite{Balitskyrc,Kovchegov-Weigertrc}.

\subsection{Running coupling Balitsky-Kovchegov equation and its analytic solution}
When one resums all powers of $\alpha_s N_f$ in the evolution kernel, the $\alpha_s N_f$ corrections modify the structure of the evolution equation. The evolution equation with running coupling corrections can be expressed as\cite{Albacete-Kovchegov}
\bea
\frac{\partial S(x_{\bot}-y_{\bot},Y)}{\partial Y} &=& \int\,d^2 z_{\bot}
  \,K^{rc}(x_{\bot},y_{\bot}, z_{\bot})
  \left[S(x_{\bot}-z_{\bot}Y)\,S(z_{\bot}-y_{\bot},Y)-S(x_{\bot}-y_{\bot},Y)\right]\nonumber \\
 &&\hspace*{0.1cm} - \alpha_{\mu}^2 \, \int d^2 z_{{\bot}1} \, d^2 z_{{\bot}2} \,
  K_{\oone} (x_{\bot}, y_{\bot} ; z_{{\bot}1}, z_{{\bot}2}) \, \left[ S (x_{\bot}-
   w_{\bot}, Y) \, S (w_{\bot}-y_{\bot}, Y)\right.\nonumber\\
 &&\hspace*{0.1cm}- \left.S(x_{\bot}-z_{{\bot}1}, Y) \, S(z_{{\bot}2}-y_{\bot}, Y)\right],
\label{BK_RCf}
\eea
where $w_{\bot}$ is the subtraction point which can be chosen to be the transverse coordinate of the emitted gluon $z_{\bot}$ or the transverse coordinate of either the quark $z_{{\bot}1}$ or the antiquark $z_{{\bot}2}$. From Eq.(\ref{BK_RCf}) we know that it has two parts, `running coupling' part (the first line on the r.h.s of Eq.(\ref{BK_RCf})) and `subtraction' part (the second line on the r.h.s of Eq.(\ref{BK_RCf})). The `running coupling' part has the same structure as the LO BK evolution equation but with a modified kernel $K^{rc}$. The `subtraction' part has a new structure with two quadratic terms and a resummed JIMWLK kernel $K_{\oone}$. It has been found that the separation between the running coupling and subtraction contributions is not unique, which depends on the choice of subtraction point\cite{Balitskyrc,Kovchegov-Weigertrc}. Fortunately, we are only interesting in the analytic solution of the evolution equation in the saturation region in which the running coupling BK equation is independent of the selection of subtraction point\cite{Xiang} since in this regime the $S$-matrix is so small that its quadratic terms in Eq.(\ref{BK_RCf}) can be neglected. Therefore the running coupling BK evolution equation simplifies to
\be
\frac{\partial S(r,Y)}{\partial Y}=-\int\,d^2 z_{\bot}
  \,K^{rc}(r, r_1, r_2)S(r, Y),
\label{SKW_sr}
\ee
where the modified kernel $K^{rc}(r, r_1, r_2)$ has two kinds of expressions (Balitsky and Kovchegov-Weigert kernels) since two different separation schemes have been used (please see Refs.\cite{Albacete-Kovchegov,Xiang} for more details about the kernels). Although Balitsky and Kovchegov-Weigert kernels have a different format at a glance, it has been shown in Ref.\cite{Xiang} that both kernels reduce to a unique form, (\ref{eq_rcBK_K_kw}), in the saturation region. In this study we adopt the choice proposed by Kovchegov-Weigert in Ref.\cite{Kovchegov-Weigertrc}, the modified kernel is written as\cite{Kovchegov-Weigertrc}
\be
  K^{rcKW}(r, r_1, r_2)=\frac{N_c}{2\pi^2}\left[
    \alpha_s(r_1^2)\frac{1}{r_1^2}-
    2\,\frac{\alpha_s(r_1^2)\,\alpha_s(r_2^2)}{\alpha_s(R^2)}\,\frac{
      {\bf r}_1\cdot {\bf r}_2}{r_1^2\,r_2^2}+
    \alpha_s(r_2^2)\frac{1}{r_2^2} \right],
\label{Krc_kw}
\ee
with
\be
R^2(r, r_1, r_2)=r_1\,r_2\left(\frac{r_2}{r_1}\right)^
{\frac{r_1^2+r_2^2}{r_1^2-r_2^2}-2\,\frac{r_1^2\,r_2^2}{
      {\bf r}_1\cdot{\bf r}_2}\frac{1}{r_1^2-r_2^2}}.
\label{R}
\ee
In the saturation regime, the dominant integral region on the r.h.s of Eq.(\ref{SKW_sr}) comes from either $1/Q_s\ll \mid\mathbf{r_1}\mid\ll \mid\mathbf{r}\mid,~\mid\mathbf{r_2}\mid\sim \mid\mathbf{r}\mid$ or $1/Q_s\ll \mid\mathbf{r_2}\mid\ll \mid\mathbf{r}\mid,~\mid\mathbf{r_1}\mid\sim \mid\mathbf{r}\mid$.
In this work we choose the first one, the modified kernel becomes
\be
K^{rcKW}({\bf r},{\bf r}_1,{\bf r}_2)=\frac{N_c}{2\pi^2} \alpha_s(r_1^2)\frac{1}{r_1^2}.
\label{eq_rcBK_K_kw}
\ee
We would like to note that if one works in the second region, the same result should be obtained. Here $\alpha_s$ is not fixed, and we use the the running coupling at one loop accuracy
\be
\alpha_s(r_1^2)=\frac{\mu}{1 + \mu_1\ln\left(\frac{1}{r_1^2\Lambda^2}\right)}.
\label{eq_rc}
\ee

Substituting the simplified kernel~(\ref{eq_rcBK_K_kw}) into~(\ref{SKW_sr}), and using (\ref{eq_rc}), we can get the running coupling BK equation in the saturation region
\be
\frac{\partial S(r,Y)}{\partial Y}=-2\frac{N_c}{2\pi^{2}}\int_{1/Q_s^2}^{r^2}\,d^2r_{1}
  \,\frac{\alpha_s(r_1^2)}{r_1^2}S(r,Y),
\label{eq S_rcBK}
\ee
whose analytic solution is\cite{Xiang}
\bea
S(r,Y) &=& \exp\left[-\frac{N_c\mu}{c\pi\mu_1}\left(\ln^2\left(\frac{Q_s^2(Y)}{\Lambda^2}\right)\ln\left(\frac{1+\mu_1\ln\frac{Q_s^2}
{\Lambda^2}}{1+\mu_1\ln\frac{1}{r^2\Lambda^2}}-\frac{1}{2}\right)+\frac{1}{\mu_1}\ln\left(\frac{Q_s^2(Y)}{\Lambda^2}\right)
\right.\right.\nn
 &&\hspace*{0.9cm}-\left.\left.\frac{1}{\mu_1^2}\ln\left(1+\mu_1\ln\frac{Q_s^2}{\Lambda^2}\right)\right)\right]S(r_0, Y_0),
\label{eq rcBK_solution}
\eea
with
\be
\ln(Q_s^2(Y)/\Lambda^2)=\sqrt{c(Y-Y_0)}+\mathcal{O}(Y^{1/6}).
\ee

It is important to stress that the rapidity dependence of the $S$-matrix in the running coupling case is different from the one in fixed coupling case. The exponent of the $S$-matrix
in the running coupling case, Eq.(\ref{eq rcBK_solution}), decreases linearly with rapidity while the exponent of the $S$-matrix has a quadratic decrease with rapidity in the fixed coupling case, see Eq.(\ref{Sol_Kovchegov}).
%-----------------------------------------------------------
\subsection{Full next-to-leading order Balitsky-Kovchegov equation and its analytic solution}
\label{subsec_NLL BK}
In the above section, we discuss the running coupling modified BK equation, which only considers the contributions from quark loops. However, except for the quark loop contributions there are gluon loop contributions to the kernel of the evolution equation. A comprehensive corrections should include both the contributions from the quark and gluon loops as well as from the tree gluon diagrams with quadratic and cubic nonlinearities\cite{Balitsky-Chirilli}. Combining all these contributions together, one can get a full NLO evolution equation
 \begin{align}
 \label{nlobk}
 %\hspace*{-0.7cm}
 \frac{\partial S(\bx-\by, Y)}{\partial Y} = &
 \frac{\abar}{2 \pi}
 \int \dif^2 \bz \,
 \frac{(\bx\minus\by)^2}{(\bx \minus\bz)^2 (\by \minus \bz)^2}\,
 \bigg\{ 1 + \frac{\abar}{4}
 \bigg[b\, \ln (\bx \minus \by)^2 \mu^2
 - b\,\frac{(\bx \minus\bz)^2 - (\by \minus\bz)^2}{(\bx \minus \by)^2}\nn
 &\times\ln \frac{(\bx \minus\bz)^2}{(\by \minus\bz)^2}
 +\frac{67}{9} - \frac{\pi^2}{3} - \frac{10 \Nf}{9 \Nc}-
 2\ln \frac{(\bx \minus\bz)^2}{(\bx \minus\by)^2} \ln \frac{(\by \minus\bz)^2}{(\bx \minus\by)^2}\bigg]
 \bigg\}\nn
 &\times\left[S(\bx-\bz, Y) S(\bz-\by, Y)- S(\bx-\by, Y) \right]
  +\, \frac{\abar^2}{8\pi^2}
 \int \frac{\dif^2 \bz\,\dif^2 \bu}{(\bu \minus \bz)^4}\nn
 &\times\bigg\{\left[\frac{ (\bx \minus \bz)^2 (\by \minus \bu)^2 + (\bx \minus\bu)^2 (\by \minus\bz)^2
- 4 (\bx \minus \by)^2 (\bu \minus \bz)^2}{(\bx \minus \bz)^2 (\by \minus \bu)^2 - (\bx \minus \bu)^2 (\by  \minus \bz)^2}
+ \frac{(\bx \minus \by)^2 (\bu \minus \bz)^2}{(\bx \minus \bz)^2 (\by  \minus \bu)^2}\right.\nn
 & \left.+ \frac{(\bx \minus \by)^4 (\bu \minus \bz)^4}{(\bx \minus \bz)^2 (\by  \minus \bu)^2((\bx \minus \bz)^2 (\by \minus \bu)^2 - (\bx \minus \bu)^2 (\by  \minus \bz)^2)}\right]
 \ln \frac{(\bx \minus \bz)^2 (\by \minus \bu)^2}{(\bx \minus \bu)^2 (\by  \minus \bz)^2} -2 \bigg\}\nn
 \vspace{0.3cm}
 &\times\left[S(\bx-\bu, Y) S(\bu-\bz, Y) S(\bz-\by, Y) - S(\bx-\bu, Y) S(\bu-\by, Y)\right]
 \nn
 &+\,\frac{\abar^2N_f}{8\pi^2N_c}\,
 \int \frac{\dif^2 \bu \,\dif^2 \bz}{(\bu \minus \bz)^4}
 \bigg[
 - \frac{(\bx \minus\bu)^2 (\by \minus\bz)^2 +
 (\bx \minus \bz)^2 (\by \minus \bu)^2
 - (\bx \minus \by)^2 (\bu \minus \bz)^2}{(\bx \minus \bz)^2 (\by \minus \bu)^2 - (\bx \minus \bu)^2 (\by  \minus \bz)^2}
  \nn
 & \hspace*{0.1cm} \times
 \ln \frac{(\bx \minus \bz)^2 (\by \minus \bu)^2}{(\bx \minus \bu)^2 (\by  \minus \bz)^2} + 2 \bigg]
 \left[S(\bx-\bz, Y) S(\bu-\by, Y)- S(\bx-\bu, Y) S(\bu-\by, Y) \right],
 \end{align}
where the $b=(11N_c-2N_f)/3Nc$ is the first coefficient of the $\beta$ function, $N_f$ is the number of flavors, and $\mu$ is the renormalization scale.
The full NLO BK equation shows two remarkable features in its structure as compared to the LO BK equation. First, the single integration term which gets a correction of order $\mathcal{O}(\bar{\alpha}^2_s)$ to the evolution kernel, has a similar structure as the LO BK equation. Second, there are two double integration terms, of order $\mathcal{O}(\bar{\alpha}^2_s)$, which only contain the non-linear $S$-matrix. The double integrations over the transverse coordinates $\bz$ and $\bu$ refer to partonic fluctuations involving two additional partons (besides the original quark and antiquark) at the time of collision. We use the notation $\mathbf{r} = \bx - \by$, $\mathbf{r}_1 = \bx - \bz$, $\mathbf{r}'_1 = \bx - \bu$, $\mathbf{r}_2=\bz - \by$, and $\mathbf{r}'_2 = \bz' - \by$ for the sizes of parent dipole and of the new daughter dipoles produced by the evolutions.

In the saturation regime, the unitarity corrections become important or $S$ is very small. Therefore, the non-linear terms can be neglected in Eq.(\ref{nlobk}). The full NLO BK equation in saturation regime becomes
\be
 \frac{\partial S(r, Y)}{\partial Y} =
 - \int \dif^2 z_{\bot} K^{fNLO}(r, r_1, r_2) S(r, Y),
\label{fnlobk}
\ee
with the modified kernel
\bea
K^{fNLO}(r, r_1, r_2) = \frac{\abar(r^2)}{2 \pi} \left\{\frac{r^2}{r_1^2r_2^2} + \frac{1}{r_1^2}\left[\frac{\alpha_s(r_1^2)}{\alpha_s(r_2^2)} - 1\right] + \frac{1}{r_2^2}\left[\frac{\alpha_s(r_2^2)}{\alpha_s(r_1^2)} - 1\right]\right.\nonumber\\
 \left.\hspace*{0.9cm} +\frac{\bar{\alpha}_s(r^2)}{4}\frac{r^2}{r_1^2r_2^2}\left[\frac{67}{9} - \frac{\pi^2}{3} - \frac{10N_f}{9N_c} - 2\ln\frac{r_1^2}{r^2}\ln\frac{r_2^2}{r^2}\right]\right\}.
\label{fnlokernel}
\eea
To analytically solve Eq.(\ref{fnlobk}), one should to work in either $1/Q_s\ll \mid\mathbf{r_1}\mid\ll \mid\mathbf{r}\mid,~\mid\mathbf{r_2}\mid\sim \mid\mathbf{r}\mid$ or $1/Q_s\ll \mid\mathbf{r_2}\mid\ll \mid\mathbf{r}\mid,~\mid\mathbf{r_1}\mid\sim \mid\mathbf{r}\mid$ region as mentioned in the LO case. If one chooses the first regime, the NLO kernel can simplify as following
\bea
K^{fNLO}(r, r_1, r_2) &=& \frac{\abar(r^2)}{2 \pi}\left[\frac{1}{r_1^2}\frac{\alpha_s(r_1^2)}{\alpha_s(r^2)} +
\frac{1}{r^2}\left(\frac{\alpha_s(r^2)}{\alpha_s(r_1^2)} - 1\right) + \frac{\bar{\alpha}_s(r^2)}{4}\frac{1}{r_1^2}\left(\frac{67}{9} - \frac{\pi^2}{3} - \frac{10N_f}{9N_c} - 2\ln\frac{r_1^2}{r^2}\ln\frac{r_2^2}{r^2}\right)\right]\nonumber\\
&\simeq& \frac{\abar(r_1^2)}{2\pi r_1^2} +
\frac{\bar{\alpha}^2_s(r^2)}{8\pi r_1^2}\left(\frac{67}{9} - \frac{\pi^2}{3} - \frac{10N_f}{9N_c}\right).
\label{kfnlos}
\eea
Substituting the simplified kernel into Eq.(\ref{fnlobk}), the evolution equation becomes
\be
 \frac{\partial S(r, Y)}{\partial Y} =
 - 2\frac{1}{2\pi}\int_{1/Q_s^2}^{r^2} \dif^2 r_1 \left[\frac{\abar(r_1^2)}{r_1^2} +
\frac{\bar{\alpha}^2_s(r^2)}{4}\frac{1}{r_1^2}\left(\frac{67}{9} - \frac{\pi^2}{3} - \frac{10N_f}{9N_c}\right)\right] S(r, Y),
 \label{nll_eq}
\ee
and has the following solution
\bea
S(r, Y) &=& \exp\left[-\frac{N_c\mu}{c\pi\mu_1}\left(\frac{2C_r}{3}\ln^3\frac{Q_s^2(Y)}
{\Lambda^2} + \ln^2\frac{Q_s^2(Y)}{\Lambda^2}\ln\frac{(r^2\Lambda^2)^{C_r} + (r^2\Lambda^2)^{C_r}\mu_1\ln\frac{Q_s^2(Y)}{\Lambda^2}}
{1 + \mu_1\ln\frac{1}{(r^2\Lambda^2)}}\right.\right.\nn
 &&\hspace*{0.9cm} + \left.\left.\frac{1}{\mu_1}\ln\frac{Q_s^2(Y)}{\Lambda^2} - \frac{1}{\mu_1^2}\ln\left(1+\mu_1\ln\frac{Q_s^2(Y)}{\Lambda^2}\right)\right)\right]S(r_0, Y_0),
 \label{nllsol}
\eea
with $\ln(Q_s^2(Y)/\Lambda^2)=\sqrt{c(Y-Y_0)}+\mathcal{O}(Y^{1/6})$ and $C_r = \alpha_s^2(r^2)N_c\mu_1(67/9 - \pi^2/3 - 10N_f/9N_c)/4\pi\mu$. From this solution we can see that the full NLO corrections (quark plus gluon loop contributions) bring a large change to the $S$-matrix. The linear rapidity dependence of the running coupling solution, Eq.(\ref{eq rcBK_solution}), is now replaced by the rapidity raised to the power of $3/2$ in the full NLO case, see Eq.(\ref{nllsol}). By comparing the solutions in LO, running coupling and full NLO cases, one can find that the running coupling effect (only quark loops) makes the quadratic rapidity ($\sim Y^2$) dependence of the exponent of $S$-matrix changing to linear ($\sim Y$) dependence, while the full NLO effects (quark plus gluon loops) let the linear $Y$ dependence returning to $Y^{3/2}$ dependence, since the gluon loop contributions compensate part of decrease.

We would like to note that the numerical solution of the full NLO BK equation show that it is unstable, the dipole amplitude decreases with growing energy and can switch to a negative value\cite{Lappi-Mantysaari}. It has been found that the instability comes from a large double-logarithmic contribution\cite{IMMST}. To solve this problem, one needs to resum double transverse logarithms to all orders and gets a double logarithmic approximation (DLA) evolution equation which can be extend to full next-to-leading logarithmic (NLL) accuracy via including the quark and gluon loop contributions. We usually call this stabilized high-order evolution equation as NLL BK equation. Our previous studies in Ref.\cite{Xiang2} found that in the saturation region the NLL BK equation has a same analytic solution as the NLO BK equation since the DLA kernel is equal to one under the saturation condition. This is why we neglect the detail derivation of the NLL BK equation and its solution in this study.

%------------------------------------------------------------------------------
\section{EFFECTS OF RARE FLUCTUATIONS IN THE CENTER OF MASS FRAME}
\label{sec:rare CM}
In high-energy region where $S$ is small, the rare fluctuation effects can play an important role in the evolution of dipole amplitude\cite{Iancu-Mueller}. These fluctuations are rare and unimportant in a general inelastic collision, but can dramatically affect the $S$-matrix in the high energy limit. The significance of the rare fluctuations can be viewed by comparing the following two scattering processes. First, if one lets two typical configurations (of condensate type) colliding each other and computes the $S$-matrix of this interaction in the center of mass frame at relative rapidity $Y$, the $S$-matrix of this condensate-condensate scattering is proportional to $\exp\{-\textit{const}~Q_s^2(Y/2) r_0^2/\alpha_s^2\}$, where $r_0$ is the size of parent dipole. Second, if one performs a scattering of a dipole (unevolved) on a typical configuration (highly evolved) and calculates the $S$-matrix of this interaction process which can be described by using the LO BK equation, one gets the $S$-matrix being proportional to $\exp\left[-\textit{const}~\ln^2(Q_s^2r_0^2)\right]$. By comparing the $S$-matrixes of the two scattering processes mentioned above, we can see that the $S$-matrix of the condensate-condensate scattering is much smaller than the one from the dipole-condensate interaction. It has been found that the inconsistency comes from the use of typical configurations to calculate the elastic scattering $S$-matrix\cite{Iancu-Mueller}. Actually, in the saturation region the $S$-matrix is dominated by rare configurations which are dilute states with few gluons.

%-----------------------------------------------------------
\subsection{Leading-order case}
\label{subsec_LO BK rare CM}
\begin{figure}[h!]
\vspace{-1.5cm}
\begin{center}
\epsfig{file=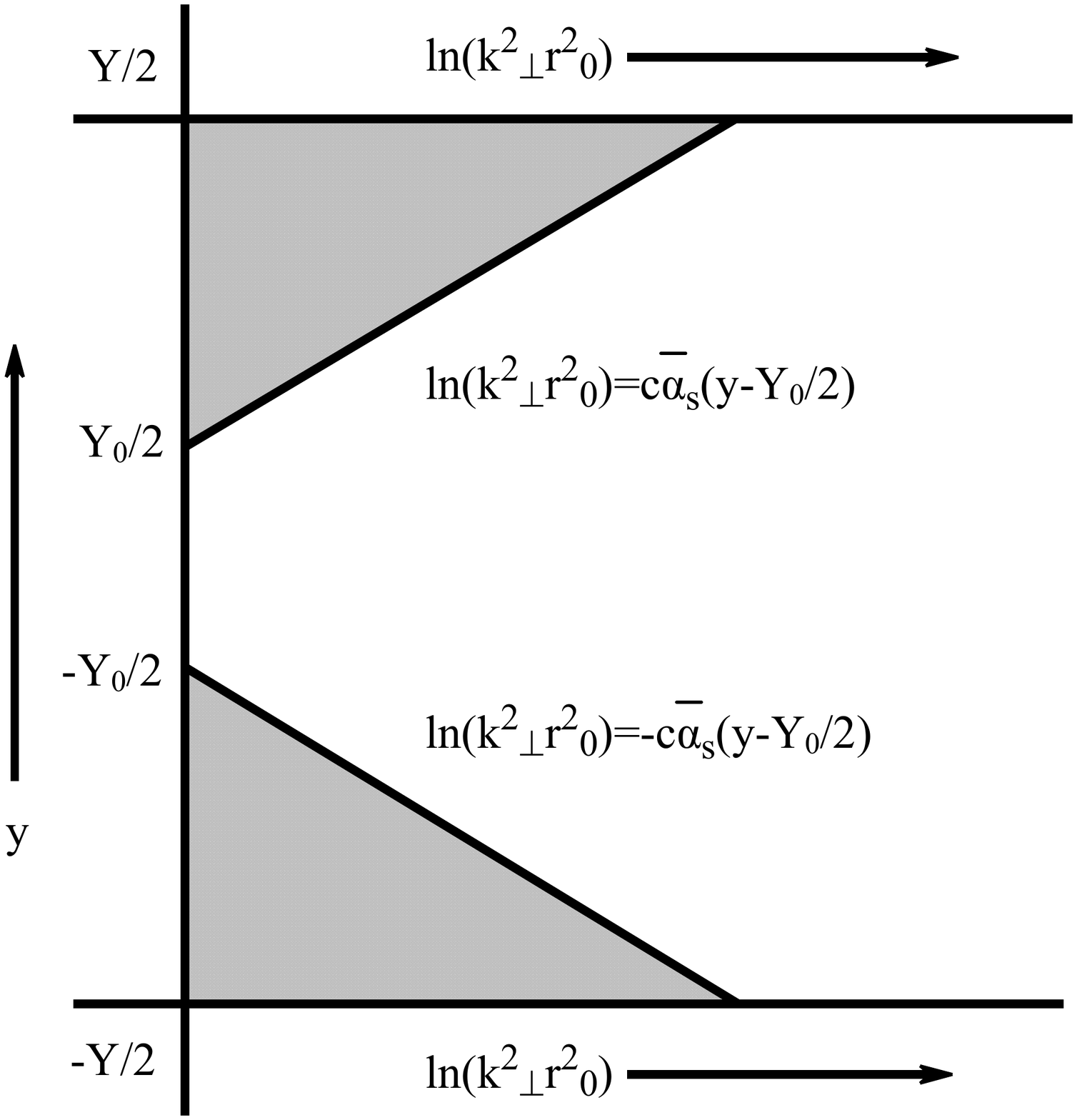, scale=0.35}
\end{center}
\vspace{-1.5cm}
\caption{The LO configuration in center of mass frame.}
\label{figLOCMF}
\end{figure}

The BK equation is obtained under a mean field approximation, which uses the product $S(r_1,Y)S(r_2,Y)$ replacing $S^{(2)}(r_1,r_2,Y)$. This replacement is only true in the absence of the fluctuations. It has been found that the $S$-matrix derived from the BK equation is too small since the wavefunctions contain too many gluons at the time of collisions\cite{Iancu-Mueller,Xiang}. In order to get a correct $S$-matrix, one needs to search for rare configurations having less than the mean number of gluons, which can lead to a larger $S$-matrix. The best way to find the rare configurations is to minimize the number of gluons by suppressing the evolution of the system. As it was done in Ref.~\cite{Iancu-Mueller}, we consider a central scattering of a right-moving dipole on a left-moving dipole in the center of mass frame at rapidity $Y > Y_{0}$, where $Y_{0}$ denotes the critical value for the onset of unitarity corrections. To obtain a rare configuration, we suppose that the wavefunction of the right-moving dipole consists only of the parent dipole with size $r_0$ in the rapidity interval $Y_0/2<y<Y/2$, but allowing normal BFKL evolution in the rapidity interval $0<y<Y_0/2$. For the left-moving dipole, we take a same requirement on the wavefunction as the right-moving dipole in the rapidity interval $-Y/2<y<-Y_0/2$ and $-Y_0/2<y<0$, respectively. Unfortunately, it is impossible to require that all evolutions are absent in the forbidden rapidity intervals. Taking right-moving dipole as an example, what one can do is to restrict that the evolution produces only very small dipoles, which can guarantee the system has no more than one dipole of size $r_0$ or larger in the rapidity interval $Y_0/2<y<Y/2$. So, it has to suppress the creation of dipoles much smaller than $r_0$ at rapidities $y>Y_0/2$ to avoid dipoles emitted at intermediate rapidities evolving into dipoles of size $r_0$ or larger at rapidity $Y/2$. As illustrated in Fig.\ref{figLOCMF}, the gluon emission from the parent dipoles is forbidden if the gluon has $k_{\bot}$ and $y$ in the shaded triangles. The line for upper triangle in Fig.\ref{figLOCMF}, $\ln(k_{\bot}^{2}r_{0}^2)=c\bar{\alpha}_s(y-Y_0/2)
\label{uplineCMF}$, is determined by the requirement that gluons on the right hand side of that line can not evolve by normal BFKL evolution into shaded triangle and the line for the lower triangle has the same requirement as the upper one.

By carrying out the operations mentioned above, the LO $S$-matrix including the rare fluctuations can be computed\cite{Iancu-Mueller}
\be
S(r,Y)=\mathbb{S}\left(r,\frac{Y-Y_0}{2}\right)\mathbb{S}\left(r,\frac{Y-Y_0}{2}\right)S(r_0,Y_0),
\label{SLOrareCM}
\ee
where $\mathbb{S}(r,(Y-Y_0)/2)$ is the probability of the rare configuration and has the same meaning as the parent dipole's survival probability after a BFKL evolution over a rapidities $Y-Y_0$\cite{Iancu-Mueller}. The terminology survival probability implies that $\mathbb{S}(r,(Y-Y_0)/2)$ satisfies the same evolution equation as the virtual term of the LO BK equation
\be
  \frac{\partial}{\partial Y}\mathbb{S}\left(r,\frac{Y-Y_0}{2}\right)=-\int d^2z_1 K^{LO}({\bf r},{\bf r}_1,{\bf r}_2)
           \mathbb{S}\left(r,\frac{Y-Y_0}{2}\right).
\label{LOrare}
\ee
Substituting the LO BK kernel into Eq.(\ref{LOrare}), one can get
\be
  \frac{\partial}{\partial y}\mathbb{S}\left(r,\frac{Y-Y_0}{2}\right)=- \int_{1/Q_s^2}^{r^2}\,d^2r_1
  \frac{2\bar{\alpha}_s}{2 \pi} \frac{r^2}{r_1^2r_2^2}\mathbb{S}\left(r,\frac{Y-Y_0}{2}\right).
\label{LOrare2}
\ee
By a simple algebra calculating, we can obtain the solution of Eq.(\ref{LOrare2})
\be
\mathbb{S}\left(r,\frac{Y-Y_0}{2}\right)=\exp\left[-\frac{c}{2}\bar{\alpha}_s^2\frac{(Y-Y_0)^2}{4})\right].
\label{LOrareres}
\ee
Taking this solution into Eq.(\ref{SLOrareCM}), one can obtain\cite{Iancu-Mueller}
\be
S(r,Y)=\exp\left[-\frac{c}{4}\bar{\alpha}_s^2(Y-Y_0)^2\right]S(r_0,Y_{0}),
\label{SLOrareCMres}
\ee
which is significantly larger than the one in Eq.(\ref{Sol_Kovchegov}) by comparing the exponent of the $S$-matrix. From Eq.(\ref{SLOrareCMres}), we can see that the exponential factor in Eq.(\ref{Sol_Kovchegov}) is twice as large as the result, Eq.(\ref{SLOrareCMres}), which includes the rare fluctuations. The rare fluctuations lead the $S$-matrix becoming larger than the one coming from LO BK equation, which indicates that the effects of rare fluctuations are very important in the LO case and cannot neglect when one studies gluon saturation phenomenology in the saturation region at fixed coupling constant.
%-----------------------------------------------------------
\subsection{Next-to-leading order case}
\label{subsec_rcBK rare CM}
\begin{figure}[h!]
\vspace{-1.5cm}
\begin{center}
\epsfig{file=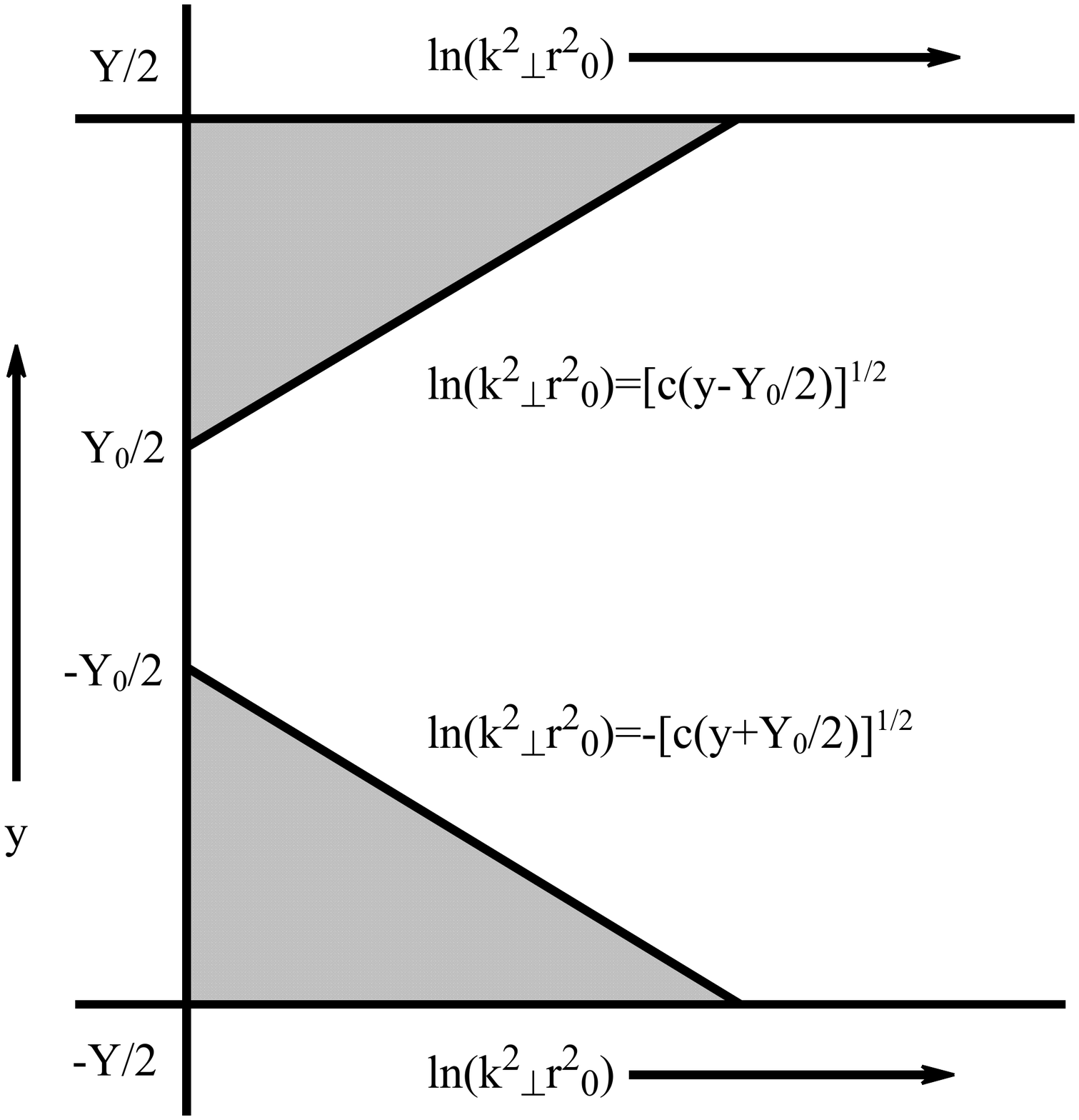, scale=0.35}
\end{center}
\vspace{-1.5cm}
\caption{The NLO configuration in center of mass frame.}
\label{figNLOCMF}
\end{figure}

In this part we investigate the rare fluctuations on top of the running coupling and full NLO effects. To get the rare configuration, we need to follow the steps in our previous studies\cite{Xiang}. Consider a right-moving dipole scattering off a left-moving dipole, one needs to constrain the wavefunctions of the right-moving and left-moving dipoles in order to let the system consisting only of the parent dipole itself with size $r_0$ in the rapidity intervals $Y_0/2<y<Y/2$ and $-Y/2<y<-Y_0/2$, respectively. However, one cannot require that all evolution of the right-moving and left-moving dipoles are absent in the rapidity intervals mentioned above. A feasible way that one can take is to allow that the evolution can only produce very small dipoles with size much smaller than $r_0$ to avoid these dipoles evolving into dipoles with similar size as $r_0$ in intermediate rapidities, which can guarantee that the final dipole system at rapidity $Y/2$ has no more than one dipole of size $r_0$ or larger, see Fig.\ref{figNLOCMF}. Note that the lines for upper and lower triangles in Fig.\ref{figNLOCMF},
$\ln(k_{\bot}^{2}r_0^2)=\sqrt{c(y-Y_0/2)}$ and $\ln(k_{\bot}^{2}r_0^2)=-\sqrt{c(y+Y_0/2)}$, are determined in the same way as the LO case except for the NLO corrections included here. So the linear rapidity dependence of the line in LO case is now replaced by square root dependence in the NLO case.
\subsubsection{In the case of running coupling}
We follow Eq.(\ref{SLOrareCM}) to derive $S$-matrix which includes both rare fluctuation and running coupling corrections. As the LO case, the probability of a rare configuration in the running coupling case satisfies
\be
  \frac{\partial}{\partial Y}\mathbb{S}(r,Y-Y_0)=-\int d^2z_1 K^{rcKW}(r, r_1, r_2)
           \mathbb{S}(r,Y-Y_0),
\label{rcBKrare}
\ee
where $K^{rcKW}(r, r_1, r_2)$  is the modified kernel which is given by Eq.(\ref{eq_rcBK_K_kw}). By solving the integro-differential equation, Eq.(\ref{rcBKrare}), one gets
\bea
\mathbb{S}(r,Y-Y_0)&=&\exp\left[-\frac{N_c\mu}{c\pi\mu_1}\left(\ln^2\left(\frac{Q_S^2(Y)}{\Lambda^2}\right)
\ln\left(\frac{1+\mu_1\ln\left(\frac{Q_S^2(Y)}{\Lambda^2}\right)}{1+\mu_1\ln\left(\frac{1}{r^2\Lambda^2}\right)}
-\frac{1}{2}\right)+\frac{\ln\left(\frac{Q_S^2(Y)}{\Lambda^2}\right)}{\mu_1}\right.\right.\nn
&&~~~~~~~-\left.\left.\frac{1}{\mu_1^2}\ln\left(1+\mu_1
\ln\left(\frac{Q_S^2(Y)}{\Lambda^2}\right)\right)\right)\right].
\label{rcBKrare_S}
\eea
By using Eq.(\ref{rcBKrare_S}), the $S$-matrix including both rare fluctuation and running coupling corrections can be computed as
\bea
S(r,Y)&=&\mathbb{S}\left(r,\frac{Y-Y_0}{2}\right)\mathbb{S}\left(r,\frac{Y-Y_0}{2}\right)S(r_0,Y_0)\nn
&=&\exp\left[-\frac{N_c\mu}{c\pi\mu_1}\left(\ln^2\left(\frac{Q_S^2(Y)}{\Lambda^2}\right)
\ln\left(\frac{1+\frac{\mu_1}{\sqrt{2}}\ln\left(\frac{Q_S^2(Y)}{\Lambda^2}\right)}
{1+\mu_1\ln\left(\frac{1}{r^2\Lambda^2}\right)}-\frac{1}{2}\right)\right.\right.\nn
&&~~~~~~~+\left.\left.\frac{\sqrt{2}
\ln\left(\frac{Q_S^2(Y)}{\Lambda^2}\right)}{\mu_1}-\frac{2}{\mu_1^2}\ln\left(1+\frac{\mu_1}
{\sqrt{2}}\ln\left(\frac{Q_S^2(Y)}{\Lambda^2}\right)\right)\right)\right]S(r_0,Y_0),
\label{SrcBKrareres}
\eea
where $\ln^2(Q_S^2/\Lambda^2)=\sqrt{c(Y-Y_0)}$.
To see whether the rare fluctuation effects bringing in a large change to the $S$-matrix or not, one can carry out a comparison between Eq.(\ref{SrcBKrareres}) and Eq.(\ref{eq rcBK_solution}), it is easy to find that in the exponent of the $S$-matrix the dominant terms are almost the same and the sub-terms have slightly modifications, which mean the rare fluctuations are less important in the running coupling case as compared to the fixed coupling case where the rare fluctuations correct the exponential factor of $S$-matrix by a factor of two. The reason why the rare fluctuations are less important is that the running coupling makes the exponent of the $S$-matrix changing to a linear rapidity dependence instead of a quadratic dependence in the fixed coupling case.

%-----------------------------------------------------------
\subsubsection{In the case of full next-to-leading order}
\label{subsec_NLO BK rare CM}
The rare configuration at the full NLO case is similar to the one in running coupling case, see Fig.\ref{figNLOCMF}. The probability of rare configuration $\mathbb{S}(r,Y-Y_0)$ satisfies the evolution equation
\be
  \frac{\partial}{\partial Y}\mathbb{S}(r,Y-Y_0)=-\int d^2z_1 K^{fNLO}({\bf r},{\bf r}_1,{\bf r}_2)\mathbb{S}(r,Y-Y_0),
\label{fnloBKrare}
\ee
whose solution is

\bea
\mathbb{S}(r,Y-Y_0) &=& \exp\left[-\frac{N_c\mu}{c\pi\mu_1}\left(\frac{2C_r}{3}(c(Y-Y_{0}))^{3/2}
 +c(Y-Y_{0})\ln\frac{(r^2\Lambda^2)^{C_r} + (r^2\Lambda^2)^{C_r}\mu_1(c(Y-Y_{0}))^{1/2}}
{1 + \mu_1\ln\frac{1}{(r^2\Lambda^2)}}\right.\right.\nn
 &&\hspace*{0.9cm} + \left.\left.\frac{1}{\mu_1}(c(Y-Y_{0}))^{1/2} - \frac{1}{\mu_1^2}\ln\left(1+\mu_1(c(Y-Y_{0}))^{1/2}\right)\right)\right],
\label{sol_nll_y-y0}
\eea
where we have used $\ln^2(Q_S^2/\Lambda^2)=\sqrt{c(Y-Y_0)}$. To get the $S$-matrix in the center of mass frame, we need to compute
\bea
\mathbb{S}\left(r, \frac{Y-Y_{0}}{2}\right) &=& \exp\left[-\frac{N_c\mu}{c\pi\mu_1}\left(\frac{2C_r}{3}\left(\frac{1}{2}\right)^{3/2}\left(c(Y-Y_{0})\right)^{3/2}
 +\frac{c}{2}(Y-Y_{0})\ln\frac{(r^2\Lambda^2)^{C_r} + (r^2\Lambda^2)^{C_r}\frac{\mu_1}{\sqrt{2}}(c(Y-Y_{0}))^{1/2}}
{1 + \mu_1\ln\frac{1}{(r^2\Lambda^2)}}\right.\right.\nn
 &&\hspace*{0.9cm} + \left.\left.\frac{1}{\sqrt{2}\mu_1}(c(Y-Y_{0}))^{1/2} - \frac{1}{\mu_1^2}\ln\left(1+\frac{\mu_1}{\sqrt{2}}(c(Y-Y_{0}))^{1/2}\right)\right)\right],
\label{sol_nll_y-yo/2}
\eea
than the $S$-matrix can be calculated as
\bea
\label{sol_nll_ycm}
S(r, Y) &=& S\left(r, \frac{Y-Y_{0}}{2}\right)S\left(r, \frac{Y-Y_{0}}{2}\right)S(r_0, Y_{0})\nn
&=& \exp\left[-\frac{N_c\mu}{c\pi\mu_1}\left(2\left(\frac{1}{2}\right)^{3/2}\frac{2C_r}{3}\ln^3\frac{Q_s^2(Y)}{\Lambda^2}
+\ln^2\frac{Q_s^2(Y)}{\Lambda^2}\ln\frac{(r^2\Lambda^2)^{C_r} + (r^2\Lambda^2)^{C_r}\frac{\mu_1}{\sqrt{2}}\ln\frac{Q_s^2(Y)}{\Lambda^2}}
{1 + \mu_1\ln\frac{1}{(r^2\Lambda^2)}}\right.\right.\nn
&&\hspace*{0.9cm} + \left.\left.\frac{\sqrt{2}}{\mu_1}\ln\frac{Q_s^2(Y)}{\Lambda^2} - \frac{2}{\mu_1^2}\ln\left(1+\frac{\mu_1}{\sqrt{2}}\ln\frac{Q_s^2(Y)}{\Lambda^2}\right)\right)\right]S(r_0,Y_0)\nn
&=& \exp\left[-\frac{N_c\mu}{c\pi\mu_1}\left(\frac{1}{\sqrt{2}}\frac{2C_r}{3}\ln^3\frac{Q_s^2(Y)}{\Lambda^2}
+\ln^2\frac{Q_s^2(Y)}{\Lambda^2}\ln\frac{(r^2\Lambda^2)^{C_r} + (r^2\Lambda^2)^{C_r}\frac{\mu_1}{\sqrt{2}}\ln\frac{Q_s^2(Y)}{\Lambda^2}}
{1 + \mu_1\ln\frac{1}{(r^2\Lambda^2)}}\right.\right.\nn
&&\hspace*{0.9cm} + \left.\left.\frac{\sqrt{2}}{\mu_1}\ln\frac{Q_s^2(Y)}{\Lambda^2} - \frac{2}{\mu_1^2}\ln\left(1+\frac{\mu_1}{\sqrt{2}}\ln\frac{Q_s^2(Y)}{\Lambda^2}\right)\right)\right]S(r_0,Y_0).
\label{SNLLrareres}
\eea
By comparing Eq.(\ref{SNLLrareres}) with Eq.(\ref{nllsol}), we can clearly see that the exponential factor of the dominant term in Eq.(\ref{nllsol}) is $\sqrt{2}$ times larger than the one which presents when the rare fluctuation corrections are taken into account. This result shows that the rare fluctuation effects become important again in the full NLO case, which does not like the case of running coupling where the rare fluctuations are washed away by the quark loop corrections. Before we explain the reason why the rare fluctuations become important again in the full NLO case, we do an analysis about the change of rapidity dependence of the $S$-matrix from LO to running coupling, and from running coupling to full NLO cases. As we obtain the solution to the LO BK equation in Eq.(\ref{Sol_Kovchegov}), the exponent has quadratic rapidity dependence, $S\sim\exp(-\mathcal{O}(Y^2))$. While this quadratic rapidity dependence is replaced by linear dependence, Eq.(\ref{eq rcBK_solution}), once the running coupling corrections are taken into account, $\exp(-\mathcal{O}(Y^2))\rightarrow \exp(-\mathcal{O}(Y))$. Further, when one includes full NLO contributions, the linear rapidity dependence of the $S$-matrix changes to rapidity raised to power of $3/2$ dependence, $\exp(-\mathcal{O}(Y))\rightarrow \exp(-\mathcal{O}(Y^{3/2}))$. The global analysis shows that the gluon loop contributions (one part of the NLO corrections) compensate part of the decrease of the rapidity made by quark loops. The compensation induces the rare fluctuations becoming important again.

%------------------------------------------------------------------------------
\section{EFFECTS OF RARE FLUCTUATIONS IN A GENERAL FRAME}
\label{sec:rare GENE}
To ensure the results what we have obtained in the above section are independent of the frame choice, we study a scattering of a right-moving dipole of size $r_0$ and rapidity $Y-Y_2$ off a left moving dipole of size $r_1$ and rapidity $-Y_2$ in a general frame. The frame and scattering pictures in the LO and NLO cases are given in Fig.\ref{figLOGEF} and Fig.\ref{figNLOGEF}, respectively.
%-----------------------------------------------------------
\subsection{Leading-order case}
\label{subsec_LO BK rare GENE}
\begin{figure}[h!]
\vspace{-1.5cm}
\begin{center}
\epsfig{file=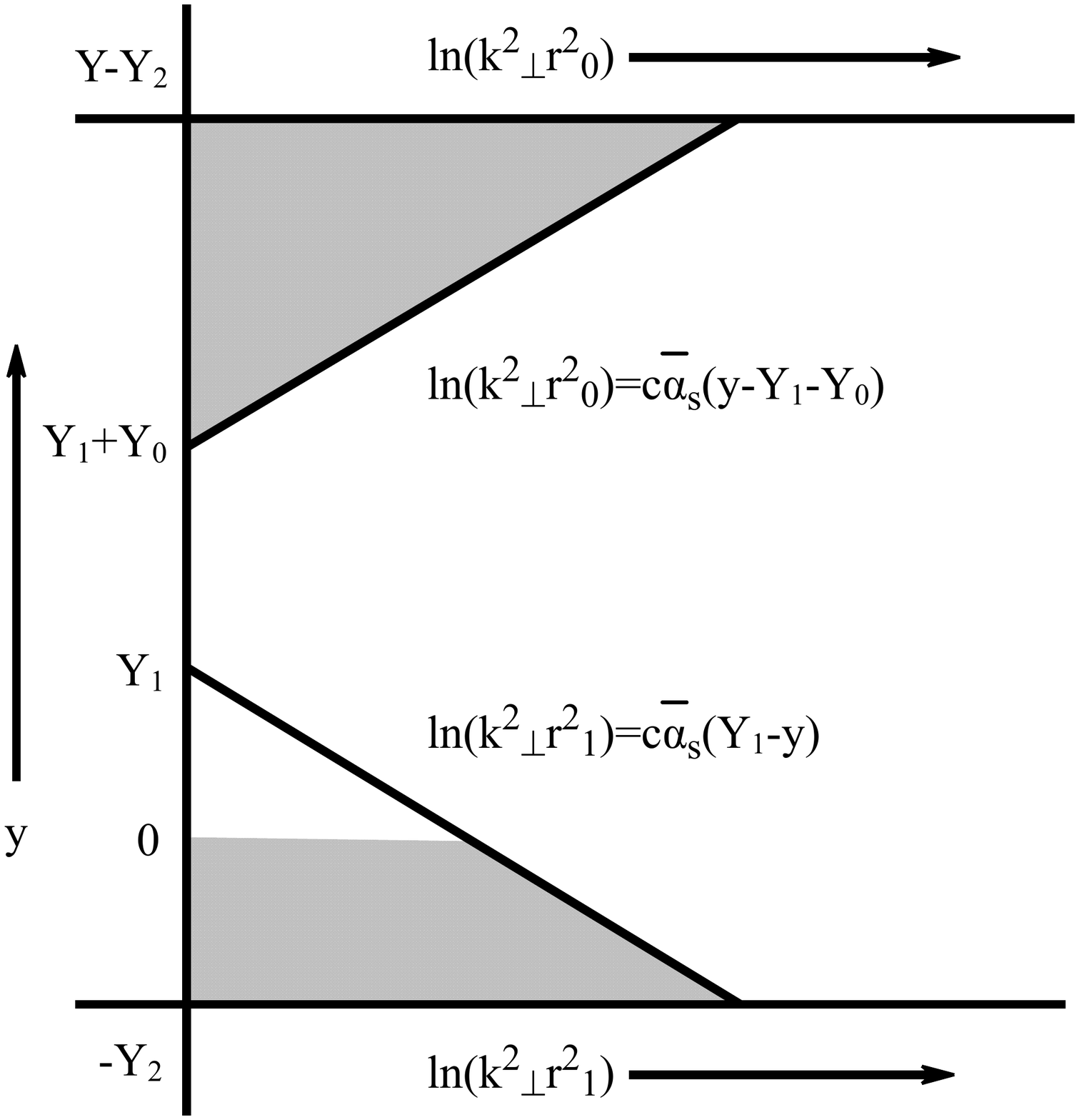, scale=0.35}
\end{center}
\vspace{-1.5cm}
\caption{The LO configuration in a general frame.}
\label{figLOGEF}
\end{figure}

We let a right-moving dipole of size
$r_0$ and rapidity $Y-Y_2$ scatter on a left-moving dipole of size $r_1$ and rapidity $-Y_2$ in a general frame. The scattering picture is illustrated in Fig.\ref{figLOGEF}. We assume that the right-moving dipole undergoes highly evolved, and the left-moving dipole has the smaller rapidity. In this study the $Y_2\leq\frac{1}{2}(Y-Y_0)$ is required for the sake of later calculation convenience, where $Y_0$ is the rapidity gap between two such dipoles in which the unitarity corrections begin to become important. In order to get a rare configuration, one has to restrict the evolution of both dipoles $r_0$ and $r_1$. For the dipole $r_0$, its evolution over the rapidity range $Y_1+Y_0<y<Y-Y_2$ (see the upper shaded triangle in Fig.\ref{figLOGEF}) is suppressed same as we did in Sec.\ref{subsec_LO BK rare CM}. In the lowest $Y_0+Y_1$ rapidity, the dipole $r_0$ has normal BFKL evolution. The unshaded triangle, $0<Y<Y_1$, is a saturation regime in which the dipole $r_0$ has evolved into a Color Glass Condensate. The $Y_1$ is an intermediate variable which will be determined by maximizing the $S$-matrix. Note that the line of the upper shaded triangle in Fig.\ref{figLOGEF}
\be
\ln(k_{\bot}^{2}r_0^2)=c\bar{\alpha}_s(y-Y_1-Y_0)
\label{uplinelogeF}
\ee
is determined by the requirement that gluons locating on the right hand side of that line can not evolve by normal BFKL evolution into the shaded triangles. We wish to note that a similar way is employed to determine the line of the lower shaded triangle in Fig.\ref{figLOGEF}.
For the dipole $r_1$, we have to strongly suppress the emission of those dipoles which can become of size $1/Q_s$ through a normal evolution over the intermediate rapidity range $-Y_2<y<0$. This means that, at the time of scattering, the left-moving dipole system has no additional dipoles of size $\lambda r_1$ or larger, with $\lambda$ a constant of order $1$.
Based on this scattering picture, we can now estimate the $S$-matrix as
\be
S(r_0,r_1,Y)=\mathbb{S}_R(r_0,Y-Y_0-Y_1-Y_2)\mathcal{S}(r_0,r_1,Y_0+Y_1)\mathbb{S}_L(r_1,Y_2),
\label{Srareloge}
\ee
where $\mathcal{S}(r_0,r_1,Y_0+Y_1)$ is the $S$-matrix for scattering of a elementary dipole $r_1$ on a Color Glass Condensate which is evolved from the dipole $r_0$. It is a dipole-typical configuration interaction and can be computed by using the LO BK equation. After using Eq.(\ref{Sol_Kovchegov}), we obtain
\be
\mathcal{S}(r_0,r_1,Y_0+Y_1)=\exp\left[-\frac{c}{2}\bar{\alpha}_s^2 Y_1^2\right]S(r_0, Y_0).
\label{Slobkrare}
\ee
As usual, the $\mathbb{S}_R(r_0,Y-Y_0-Y_1-Y_2)$ and $\mathbb{S}_L(r_1,Y_2)$  are the suppression factors which denote no emission from the two
dipoles, and are given in terms of the suppressions over the upper and
lower shaded regions in Fig.\ref{figLOGEF},
\be
\mathbb{S}_R(r_0,Y-Y_2-Y_1-Y_0)=\exp\left[-\frac{c}{2}\bar{\alpha}_s^2(Y-Y_2-Y_1-Y_0)^2\right],
\label{srlo}
\ee
and
\be
\mathbb{S}_L(r_1,Y_2)=\exp\left[-\frac{c}{2}\bar{\alpha}_s^2((Y_1+Y_2)^2-Y_1^2)\right].
\label{sllo}
\ee
Substituting (\ref{Slobkrare}), (\ref{srlo}) and (\ref{sllo}) into (\ref{Srareloge}), one obtains
\be
S(r_0,r_1,Y)=\exp\left\{-\frac{c}{2}\bar{\alpha}_s^2\left[(Y-Y_2-Y_1-Y_0)^2+(Y_1+Y_2)^2\right]\right\}S(r_0,Y_{0}).
\label{Srareloge_11}
\ee
Note that the $Y_1$ is a rapidity which describes the amount
of evolution in the right-moving dipole, and can be determined by maximizing the $S$-matrix through minimizing the exponent of Eq.(\ref{Srareloge_11}). So, one can get the optimal value of $Y_1$
\be
Y_1=\frac{1}{2}(Y-Y_0)-Y_2.
\label{Y1}
\ee
Taking $Y_1$ into Eq.(\ref{Srareloge_11}), one obtains the $S$-matrix
\be
S(r_0, r_1, Y)=\exp\left[-\frac{c}{4}\bar{\alpha}_s^2(Y-Y_0)^2\right]S(r_0,Y_{0}).
\label{SLOraregeres_gen}
\ee
As expected, this result is exactly the same as the corresponding result~(\ref{SLOrareCMres}) in the center of mass frame. We would like to point out that the right hand size of Eq.(\ref{SLOraregeres_gen}) depends on $r_1$ through the rapidity $Y_0$.

%-----------------------------------------------------------
\subsection{Next-to-leading order case}
\label{subsec_rcBK rare GENE}
\begin{figure}[h!]
\vspace{-1.5cm}
\begin{center}
\epsfig{file=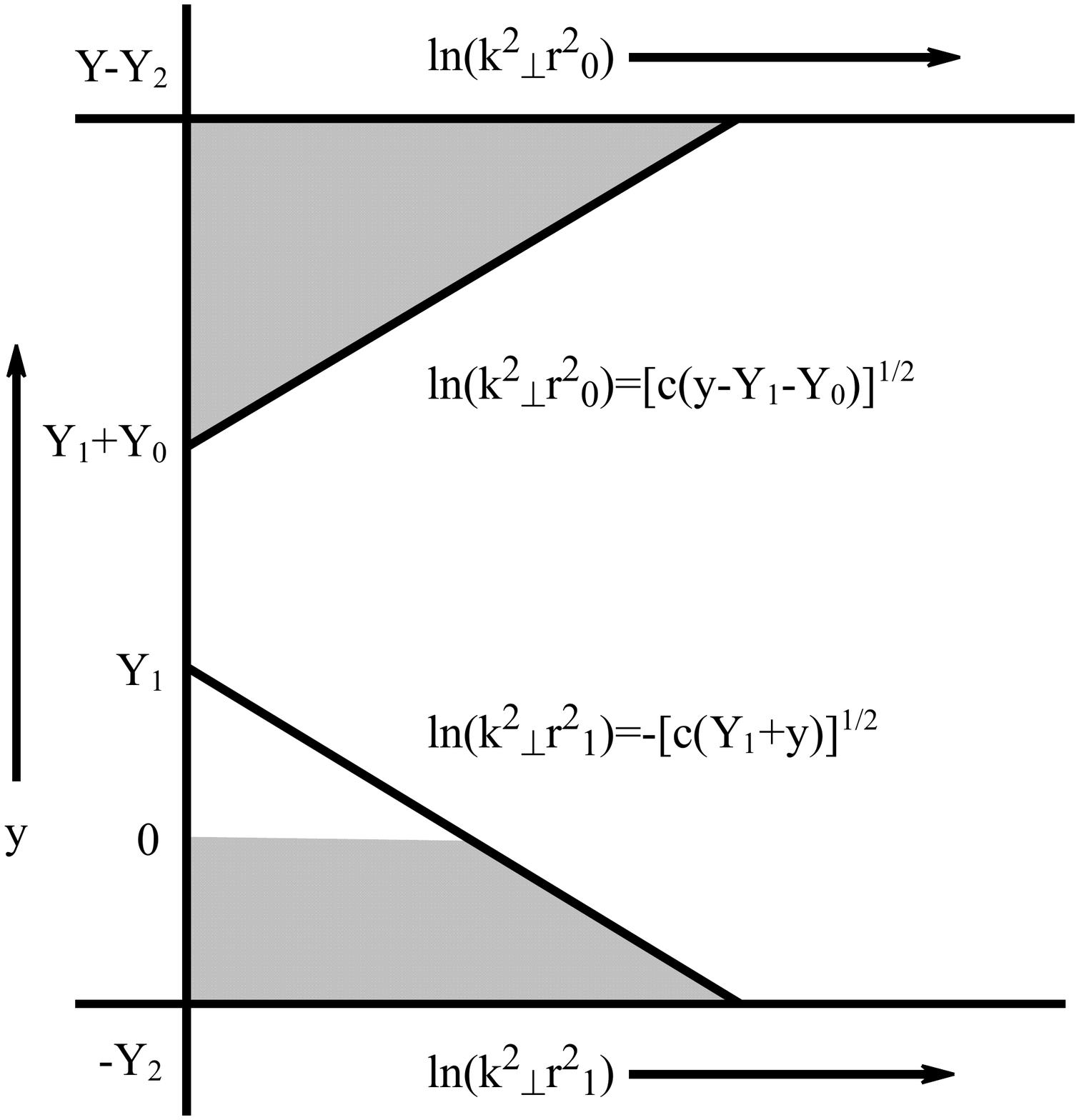, scale=0.35}
\end{center}
\vspace{-1.5cm}
\caption{The NLO configuration in a general frame.}
\label{figNLOGEF}
\end{figure}

In the next-to-leading order case, the scattering picture in a general frame is illustrated in Fig.\ref{figNLOGEF}.
The line of the upper shaded triangle in Fig.\ref{figNLOGEF}
\be
\ln(k_{\bot}^{2}r_0^2)=\sqrt{c(y-Y_1-Y_0)}
\label{uplinelogeF}
\ee
is determined by the requirement that gluons locating on the right hand side of that line can not evolve by normal BFKL evolution into the shaded triangles. We wish to note that a similar way is employed to determine the line of the lower shaded triangle in Fig.\ref{figNLOGEF}.

\subsubsection{In the case of running coupling}
We use the same way as mentioned above to compute the $S$-matrix. Based on Eq.(\ref{Srareloge}) we know that the suppression factors $\mathbb{S}_R$ and $\mathbb{S}_L$ need to be computed. Since the suppression process has similar meaning as the survival probability of the parent dipoles in a normal BFKL evolution, it can be calculated by using the virtual term in (\ref{BK_RCf}), one obtains
\bea
\mathbb{S}_R(r_0,Y-Y_0-Y_1-Y_2)&=& \exp\left[-\frac{N_c\mu}{c\pi\mu_1}\left(c\ln\left(\frac{1+\mu_1\sqrt{c(Y-Y_2-Y_1-Y_0)}}
{1+\mu_1\ln\left(\frac{1}{r^2\Lambda^2}\right)}-\frac{1}{2}\right)(Y-Y_2-Y_1-Y_0)\right.\right.\nonumber \\
&&\hspace*{0.9cm} + \left.\left.\frac{\sqrt{c(Y-Y_2-Y_1-Y_0)}}{\mu_1}- \frac{1}{\mu_1^2}\ln\left(1+\mu_1\sqrt{c(Y-Y_2-Y_1-Y_0)}\right)\right)\right],
\label{srrcBK}
\eea
and
\bea
\mathbb{S}_L(r_1,Y_2)&=&\exp\left\{-\frac{N_c\mu}{c\pi\mu_1}\left[c\ln\left(\frac{1+\mu_1\sqrt{c(Y_1+Y_2)}}
{1+\mu_1\ln\left(\frac{1}{r^2\Lambda^2}\right)}-\frac{1}{2}\right)(Y_1+Y_2)+\frac{\sqrt{c(Y_1+Y_2)}}
{\mu_1}-\frac{1}{\mu_1^2}\ln\left(1+\mu_1\sqrt{c(Y_1+Y_2)}\right) \right.\right.\nonumber \\
&&\hspace*{0.9cm} -\left.\left.c\ln\left(\frac{1+\mu_1\sqrt{cY_1}}{1+\mu_1\ln\left(\frac{1}{r^2\Lambda^2}\right)}
-\frac{1}{2}\right)Y_1-\frac{\sqrt{cY_1}}{\mu_1}+\frac{1}{\mu_1^2}\ln\left(1+\mu_1\sqrt{cY_1}\right)\right]\right\}
\label{slrcBK}
\eea
The $S$-matrix for the dipole-typical configuration interaction in the running coupling case is derived in Eq.(\ref{eq rcBK_solution}). By using Eq.(\ref{eq rcBK_solution}), the $\mathcal{S}(r_0,r_1,Y_0+Y_1)$ in Eq.(\ref{Srareloge}) can be written as
\be
\mathcal{S}(r_0,r_1,Y_0+Y_1)=\exp\left\{-\frac{N_c\mu}{c\pi\mu_1}\left[c\ln\left(\frac{1+\mu_1\sqrt{cY_1}}
{1+\mu_1\ln\left(\frac{1}{r^2\Lambda^2}\right)}-\frac{1}{2}\right)Y_1+\frac{\sqrt{cY_1}}{\mu_1}-\frac{1}
{\mu_1^2}\ln\left(1+\mu_1\sqrt{cY_1}\right)\right]\right\}S(r_0,Y_0).
\label{Srcbkrare}
\ee
Combining (\ref{srrcBK}), (\ref{slrcBK}) and (\ref{Srcbkrare}), one gets
\bea
S(r,Y)&=&\mathbb{S}_R(r_0,Y-Y_0-Y_1-Y_2)\mathcal{S}(r_0,r_1,Y_0+Y_1)\mathbb{S}_L(r_1,Y_2)\nonumber \\
&=&\exp\left\{-\frac{N_c\mu}{c\pi\mu_1}\left[c\ln\left(\frac{1+\mu_1\sqrt{c(Y-Y_2-Y_1-Y_0)}}
{1+\mu_1\ln\left(\frac{1}{r^2\Lambda^2}\right)}-\frac{1}{2}\right)(Y-Y_2-Y_1-Y_0)+\frac{\sqrt{c(Y-Y_2-Y_1-Y_0)}}{\mu_1}\right.\right.\nonumber\\
&&\hspace*{0.9cm} -\left.\left.\frac{1}{\mu_1^2}\ln\left(1+\mu_1\sqrt{c(Y-Y_2-Y_1-Y_0)}\right)
+c\ln\left(\frac{1+\mu_1\sqrt{c(Y_1+Y_2)}}{1+\mu_1\ln\left(\frac{1}{r^2\Lambda^2}\right)}-\frac{1}{2}\right)
(Y_1+Y_2)+\frac{\sqrt{c(Y_1+Y_2)}}{\mu_1}\right.\right.\nonumber\\
&&\hspace*{0.9cm} -\left.\left.\frac{1}{\mu_1^2}\ln\left(1+\mu_1\sqrt{c(Y_1+Y_2)}\right)\right]\right\}S(r_0,Y_0).
\label{SrcBKrarege}
\eea
The maximum value of the $S$-matrix can be determined by using the same way as in the leading order case. One can get the optimal value of $Y_1$ same as (\ref{Y1}). Substituting (\ref{Y1}) into (\ref{SrcBKrarege}), the final $S$-matrix is
\bea
S(r,Y)&=&\exp\left\{-\frac{N_c\mu}{c\pi\mu_1}\left[\ln^2\left(\frac{Q_S^2(Y)}{\Lambda^2}\right)\ln\left(\frac{1+\frac{\mu_1}{\sqrt{2}}
\ln\left(\frac{Q_S^2(Y)}{\Lambda^2}\right)}{1+\mu_1\ln\left(\frac{1}{r^2\Lambda^2}\right)}-\frac{1}{2}\right)\right.\right.\nonumber\\
&&\hspace*{0.9cm} +\left.\left.\frac{\sqrt{2} \ln\left(\frac{Q_S^2(Y)}{\Lambda^2}\right)}{\mu_1}-\frac{2}{\mu_1^2}\ln\left(1+\frac{\mu_1}{\sqrt{2}}\ln\left(\frac{Q_S^2(Y)}
{\Lambda^2}\right)\right)\right]\right\},
\label{arbf}
\eea
which is exactly the same as the one, Eq.(\ref{SrcBKrareres}), in the center of mass frame.

%-----------------------------------------------------------
\subsubsection{In the case of full next-to-leading order}
\label{subsec_LO BK rare GENE}
Since we have already derived the effect of rare fluctuations on top of running coupling, it is now easy to transform the corresponding formalisms to full next-to-leading case by only changing the corresponding running coupling $S$-matrix to the $S$-matrix of full next-to-leading corrections. Using Eq.(\ref{nllsol}), we obtains
\bea
\mathbb{S}_{R}(r, Y-Y_{0}-Y_{1}-Y_{2}) &=& \exp\left[-\frac{N_c\mu}{c\pi\mu_1}\left(\frac{2C_r}{3}(c(Y-Y_{0}-Y_{1}-Y_{2}))^{3/2}
 +c(Y-Y_{0}-Y_{1}-Y_{2})\right.\right.\nn
&&\hspace*{0.9cm} \times  \ln\frac{(r^2\Lambda^2)^{C_r} + (r^2\Lambda^2)^{C_r}\mu_1(c(Y-Y_{0}-Y_{1}-Y_{2}))^{1/2}}
{1 + \mu_1\ln\frac{1}{(r^2\Lambda^2)}} + \frac{1}{\mu_1}(c(Y-Y_{0}-Y_{1}-Y_{2}))^{1/2} \nn
&&\hspace*{0.9cm} - \left.\left.\frac{1}{\mu_1^2}\ln\left(1 + \mu_1(c(Y-Y_{0}-Y_{1}-Y_{2}))^{1/2}\right)\right)\right],
 \label{sol_nll_ger}
\eea
and
\bea
\mathbb{S}_{L}(r_{1}, Y_{2}) &=& \exp\left[-\frac{N_c\mu}{c\pi\mu_1}\left(\frac{2C_r}{3}(c(Y_{1}+Y_{2}))^{3/2} +c(Y_{1}+Y_{2})\right.\right.\nn
&&\hspace*{0.9cm} \times \ln\frac{(r^2\Lambda^2)^{C_r} + (r^2\Lambda^2)^{C_r}\mu_1(c(Y_{1}+Y_{2}))^{1/2}}
{1 + \mu_1\ln\frac{1}{(r^2\Lambda^2)}} + \frac{1}{\mu_1}(c(Y_{1}+Y_{2}))^{1/2} \nn
&&\hspace*{0.9cm} - \frac{1}{\mu_1^2}\ln\left(1 + \mu_1(c(Y_{1}+Y_{2}))^{1/2}\right)-\frac{2C_r}{3}(cY_{1})^{3/2}\nn
&&\hspace*{0.9cm} -cY_{1}\ln\frac{(r^2\Lambda^2)^{C_r} + (r^2\Lambda^2)^{C_r}\mu_1(cY_{1})^{1/2}}
{1 + \mu_1\ln\frac{1}{(r^2\Lambda^2)}}\nn
&&\hspace*{0.9cm} \left.\left.-\frac{1}{\mu_1}(cY_{1})^{1/2} +\frac{1}{\mu_1^2}\ln\left(1 + \mu_1(cY_{1})^{1/2}\right)\right)\right].
\label{sol_nll_gel}
\eea
Similarly, the $S$-matrix for scattering of a elementary dipole on a Color Glass Condensate state is
\bea
S(r,r_{1}, Y_{0}+Y_{1}) &=& \exp\left[-\frac{N_c\mu}{c\pi\mu_1}\left(\frac{2C_r}{3}(cY_{1})^{3/2}
+cY_{1} \ln\frac{(r^2\Lambda^2)^{C_r} + (r^2\Lambda^2)^{C_r}\mu_1(cY_{1})^{1/2}}
{1 + \mu_1\ln\frac{1}{(r^2\Lambda^2)}} \right.\right.\nn
&&\hspace*{0.9cm} \left.\left.+ \frac{1}{\mu_1}(cY_{1})^{1/2}- \frac{1}{\mu_1^2}\ln\left(1 + \mu_1(cY_{1})^{1/2}\right)\right)\right]S(r,Y_0).
 \label{sol_nll_gey0y1}
\eea
Thus,
\bea
\label{sol_nll_geres1}
S(r, Y) &=& S_{R}(r, Y-Y_{0}-Y_{1}-Y_{2})S(r,r_{1}, Y_{0}+Y_{1})S_{L}(r_{1}, Y_{2})\nn
&=& \exp\left[-\frac{N_c\mu}{c\pi\mu_1}\left(\frac{2C_r}{3}(c(Y-Y_{0}-Y_{1}-Y_{2}))^{3/2}
 +c(Y-Y_{0}-Y_{1}-Y_{2})\right.\right.\nn
&&\hspace*{0.9cm} \times  \ln\frac{(r^2\Lambda^2)^{C_r} + (r^2\Lambda^2)^{C_r}\mu_1(c(Y-Y_{0}-Y_{1}-Y_{2}))^{1/2}}
{1 + \mu_1\ln\frac{1}{(r^2\Lambda^2)}} + \frac{1}{\mu_1}(c(Y-Y_{0}-Y_{1}-Y_{2}))^{1/2} \nn
&&\hspace*{0.9cm} - \left.\frac{1}{\mu_1^2}\ln\left(1 + \mu_1(c(Y-Y_{0}-Y_{1}-Y_{2}))^{1/2}\right)\right)
-\frac{N_c\mu}{2c\pi\mu_1}\left(\frac{2C_r}{3}(c(Y_{1}+Y_{2}))^{3/2} +c(Y_{1}+Y_{2})\right.\nn
&&\hspace*{0.9cm}  \times  \ln\frac{(r^2\Lambda^2)^{C_r} + (r^2\Lambda^2)^{C_r}\mu_1(c(Y_{1}+Y_{2}))^{1/2}}
{1 + \mu_1\ln\frac{1}{(r^2\Lambda^2)}} + \frac{1}{\mu_1}(c(Y_{1}+Y_{2}))^{1/2} \nn
&&\hspace*{0.9cm} - \left.\frac{1}{\mu_1^2}\ln\left(1 + \mu_1(c(Y_{1}+Y_{2}))^{1/2}\right)\right].
\eea
Substituting the optimal value of $Y_1$ into (\ref{sol_nll_geres1}), the maximum value of $S$-matrix is
\bea
\label{sol_nll_geres1_f}
S(r, Y) &=& \exp\left[-\frac{N_c\mu}{2c\pi\mu_1}\left(\frac{2C_r}{3}2\left(c\frac{Y-Y_{0}}{2}\right)^{3/2}
 +2c\frac{(Y-Y_{0})}{2}\right.\right.\nn
&&\hspace*{0.9cm}\times  \ln\frac{(r^2\Lambda^2)^{C_r} + (r^2\Lambda^2)^{C_r}\mu_1\left(c\frac{Y-Y_{0}}{2}\right)^{1/2}}
{1 + \mu_1\ln\frac{1}{(r^2\Lambda^2)}} + \frac{2}{\mu_1}\left(c\frac{Y-Y_{0}}{2}\right)^{1/2} \nn
&&\hspace*{0.9cm} - \left.\left.\frac{2}{\mu_1^2}\ln\left(1 + \mu_1\left(c\frac{Y-Y_{0}}{2}\right)^{1/2}\right)\right)\right]\nn
&=& \exp\left[-\frac{N_c\mu}{2c\pi\mu_1}\left(\frac{1}{\sqrt{2}}\frac{2C_r}{3}\ln^3\frac{Q_s^2(Y)}{\Lambda^2}
+\ln^2\frac{Q_s^2(Y)}{\Lambda^2}\ln\frac{(r^2\Lambda^2)^{C_r} + (r^2\Lambda^2)^{C_r}\frac{\mu_1}{\sqrt{2}}\ln\frac{Q_s^2(Y)}{\Lambda^2}}
{1 + \mu_1\ln\frac{1}{(r^2\Lambda^2)}}\right.\right.\nn
&&\hspace*{0.9cm} + \left.\left.\frac{\sqrt{2}}{\mu_1}\ln\frac{Q_s^2(Y)}{\Lambda^2} - \frac{2}{\mu_1^2}\ln\left(1+\frac{\mu_1}{\sqrt{2}}\ln\frac{Q_s^2(Y)}{\Lambda^2}\right)\right)\right]S(r,Y_0).
\eea
which is exactly the same as the corresponding result~(\ref{SNLLrareres}) in the center of mass frame.

Through the calculations what we have done above, we know that the results Eqs.(\ref{SLOrareCMres}), (\ref{SrcBKrareres}) and (\ref{SNLLrareres}) are independent of the frame choice. The exponential factors of the $S$-matrixes are twice and $\sqrt{2}$ as large as the results which emerge when the rare fluctuation effects are taken into account in the LO and full NLO cases, respectively. These findings indicate that
the rare fluctuation effects are important in the LO and full NLO cases.

%-----------------------------------------------------------------------------

\begin{acknowledgments}
This work is supported by the National Natural Science Foundation of China under Grant Nos.11765005, 11305040, 11847152; the Fund of Science and Technology Department of Guizhou Province under Grant No.[2018]1023; and the Education Department of Guizhou Province under Grant No.KY[2017]004.
\end{acknowledgments}

%-----------------------------------------------------------------------------
%

%-----------------------------------------------------------------------------

\end{document}